\documentclass{ws-p8-50x6-00}

\input{psfig}

\def\Mpc{\, h^{-1} \, {\rm Mpc}}

\def\Mpc{\ifmmode {\, h^{-1} \, {\rm Mpc}}
\else {$h^{-1}\,$ Mpc}\fi}

\def\s8{{\sigma_8}}

\def\ltsima{$\; \buildrel < \over \sim \;$}
\def\simlt{\lower.5ex\hbox{\ltsima}} 
\def\gtsima{$\; \buildrel > \over \sim \;$} 
\def\simgt{\lower.5ex\hbox{\gtsima}}

\def\omegam{{\Omega_{\rm m}}}
 
\def\omegab{{\Omega_{\rm b}}}

\def\qrms{{Q_{\rm rms}}} 
\def\omegabh2{{\omegab h^2}} 

\begin{document}

\title{New Cosmological Data and the `Best-Fit' FRW Universe}

\author{Ofer Lahav}

\address{Institute of Astronomy, Madingley Road, Cambridge CB3 0HA,
UK\\E-mail: lahav@ast.cam.ac.uk}


\maketitle

\abstracts{
Observational tests for the homogeneity of the Universe on large scales
are reviewed. Assuming the Cosmological Principle we then estimate 
cosmological parameters by joint analysis of 
the Cosmic Microwave Background, Supernovae Ia, peculiar velocities, 
cluster abundance and redshift surveys.
Our results are consistent with results obtained
by other groups,  
suggesting a best-fit FRW Universe with  
$\omegam = 1 -\lambda \approx  0.3$ and $H_0 \approx 75$ km/sec/Mpc.
We point out some potential problems with this currently popular model. 
 }

\section{Introduction}

The Cosmological Principle was first adopted when observational
cosmology was in its infancy; it was then little more than a
conjecture, embodying 'Occam's razor' for the simplest possible
model. Observations could not then probe to significant redshifts, the
`dark matter' problem was not well-established and the Cosmic Microwave
Background (CMB) and the X-Ray Background (XRB)  were still unknown.  
If the  Cosmological Principle turned out to be invalid 
then the consequences to our understanding of cosmology would be dramatic, 
for example the conventional way of interpreting the age of the Universe, 
its geometry and matter content would have to be revised. 
Therefore it is 
important to revisit this underlying assumption in the light of new
galaxy surveys and measurements of the background radiations.

Like with any other idea about the physical world, we cannot 
prove a model, but only falsify it.
Proving  the homogeneity of the Universe is in particular difficult 
as we observe the Universe from one point in space, and we can only 
deduce isotropy directly.
The practical methodology we adopt is to assume homogeneity and to assess
the level of fluctuations relative to the mean, and hence to test
for consistency with  the underlying hypothesis.
If the assumption of homogeneity turns out to be wrong, then 
there are  numerous possibilities
 for inhomogeneous models, and each of them must be
tested against the observations.

Here we examine the degree of smoothness with scale 
by considering 
redshift and peculiar velocities surveys, 
radio-sources, the XRB, the Ly-$\alpha$ forest,
and the CMB.
We discuss some  inhomogeneous models
and show that a fractal model on large scales is highly improbable.
Assuming an FRW metric we evaluate the `best fit Universe' by 
performing a joint 
analysis of cosmic probes.

\section{Cosmological Principle(s)}

Cosmological Principles were stated over different periods in 
human history based on philosophical and aesthetic considerations 
rather than on fundamental physical laws.
Rudnicki (1995) summarized some of these principles in modern-day
language:

$\bullet$ The Ancient Indian: 
{\it The Universe is infinite in space and time and is 
infinitely heterogeneous}.

$\bullet$ The Ancient Greek:
{\it Our Earth is the natural centre of the Universe}.

$\bullet$ The Copernican CP:
{\it 
The Universe as 
observed from any planet looks much the same}.

$\bullet$ The Generalized CP:
{\it 
The Universe is (roughly) homogeneous and isotropic}.

$\bullet$ The Perfect CP:
{\it The Universe is (roughly) homogeneous in space and time,
and is isotropic in space}.

$\bullet$ The Anthropic Principle:
{\it A human being, as he/she is, can exist only in the Universe
as it is.}

\bigskip

We note that the Ancient Indian principle can be viewed as 
a `fractal model'. 
The Perfect CP led to the steady state model, 
which although more symmetric than the CP, 
was rejected on observational grounds.
The Anthropic Principle is  becoming popular again, e.g. in 
`explaining' a non-zero cosmological constant.
Our goal here is to quantify `roughly' in the definition of the 
generalized CP, and to assess if one may assume safely
 the Friedmann-Robertson-Walker
(FRW) metric of space-time.

\section {Probes of Smoothness}

\subsection {The CMB}  

The CMB is the strongest evidence for homogeneity.
Ehlers, Garen and Sachs (1968) showed that by combining the 
CMB isotropy with the Copernican principle 
one can deduce homogeneity. More formally the 
EGS theorem (based on Liouville theorem) states that
``If the fundamental observers in a dust spacetime see an isotropic
radiation field, then the spacetime is locally FRW''.
The COBE measurements of temperature 
fluctuations  $\Delta T/T = 10^{-5} $ on scales of $10^\circ$ give 
via the Sachs Wolfe effect ($\Delta T/T = \frac {1}{3} \Delta \phi/c^2$) 
and Poisson equation
rms density fluctuations of ${{\delta \rho} \over {\rho}} \sim 10^{-4} $ on $1000 \Mpc$ (e.g. Wu,  Lahav \& Rees 
1999; see Fig 3 here), i.e. the deviations from a smooth Universe are tiny.


\subsection {Galaxy Redshift Surveys}

 Figure 1 shows the distribution of galaxies in the ORS and IRAS 
 redshift surveys. It is apparent that the distribution is highly 
 clumpy, with the Supergalactic Plane seen in full glory.
 However, deeper surveys such as LCRS and 2dFGRS (see below) 
 show that the fluctuations 
 decline as the length-scales increase. Peebles (1993) has shown 
 that the angular correlation functions for the Lick and APM surveys 
 scale with magnitude as expected in a universe which approaches 
 homogeneity on large scales.

\begin{figure}
\protect\centerline{
\psfig{figure=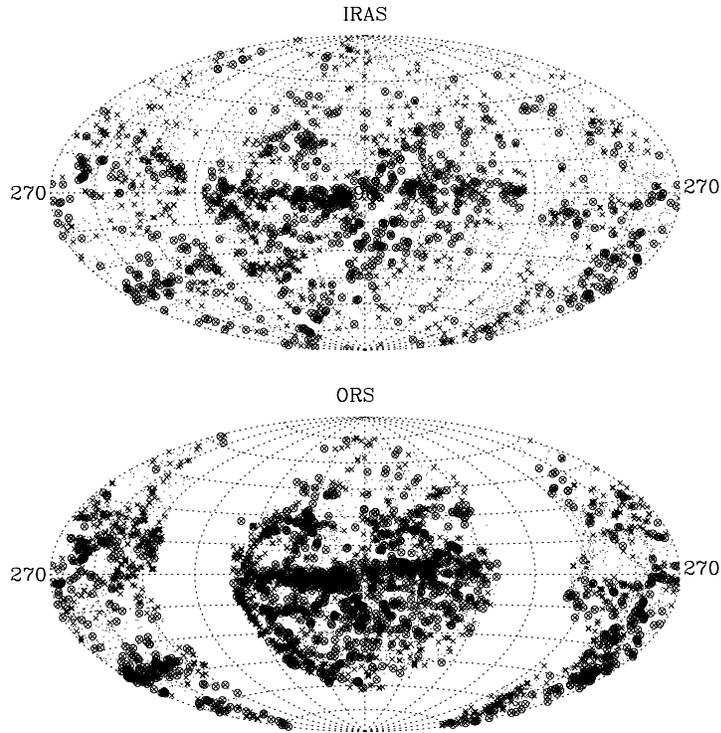,height=4truein,width=4truein}}
\caption[]{
The distribution of galaxies projected on the sky in
the IRAS and ORS samples.  This is an Aitoff projection in
Supergalactic coordinates, with $L = 90^\circ, B = 0^\circ$ (close to
the Virgo cluster) in the centre of the map. Galaxies within 2000 km/sec
are shown as circled crosses; galaxies between 2000 and 4000 km/sec are
indicated as crosses, and dots mark the positions of more distant
objects.  Here we include only catalogued galaxies, which is why the
Zone of Avoidance is so prominent in these two figures.
(Plot by M. Strauss, from Lahav et al. 2000).  }
\end{figure}

\begin{figure}
\protect\centerline{
\psfig{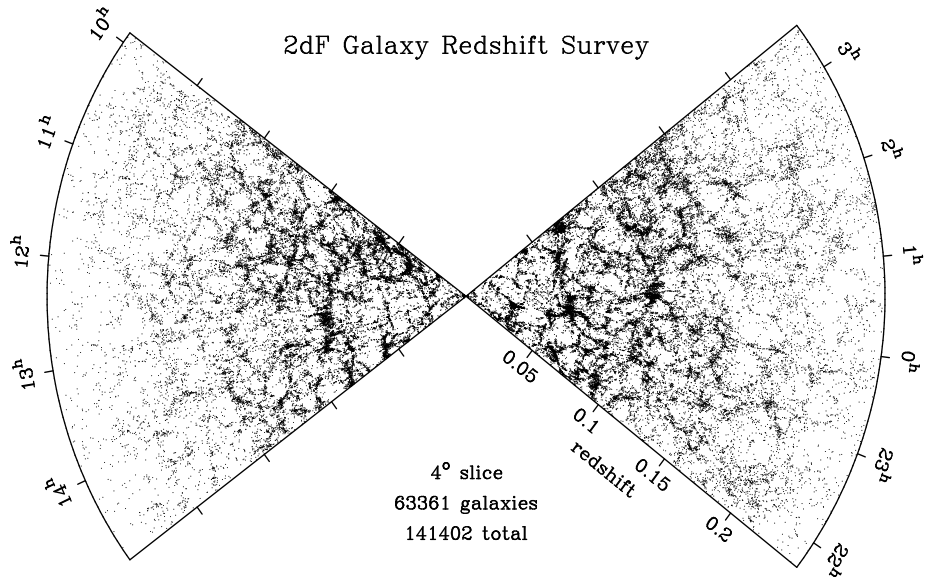}}
\caption[]{
The distribution of galaxies in part of the 2dFGRS, 
drawn from a total of 141,402 galaxies. 
The slices are 4 deg thick, towards the Northern Galactic Pole (left) 
and towards the Southern Galactic Pole (right). 
Not all 2dF fields within the slice have been observed at this stage; 
hence there are weak variations of the density of sampling. 
The image reveals a wealth of structure, 
including superclusters and voids,
but the similarity between the two slices suggests that on large scales
the universe is isotropic and homogeneous. 
(from Peacock et al. 2001).  }
\end{figure}

Multifibre technology now allows us to measure  redshifts
of millions of galaxies.  Two major surveys are underway.
The US Sloan Digital Sky Survey (SDSS) will measure redshifts to about
1 million galaxies over a quarter of the sky.  The Anglo-Australian 2
degree Field Galaxy Redshift Survey (2dFGRS) survey will measure redshifts for 250,000 galaxies
selected from the APM catalogue.  About 150,000 2dF redshifts have been
measured so far (as of March 2001).  The median redshift of both the
SDSS and the 2dFGRS galaxy redshift surveys is ${\bar z} \sim 0.1$.  While
they can provide interesting estimates of the fluctuations on 
intermediate scales (e.g. Peacock et al. 2001; see Fig 2),
the problems of biasing, evolution and
$K$-correction, would limit the ability of SDSS and 2dF to `prove' the
Cosmological Principle.  (cf. the analysis of the ESO slice by
Scaramella et al 1998 and Joyce et al. 1999).

\subsection{Peculiar Velocities}

Peculiar velocities are powerful as they probe directly the mass distribution
(e.g. Dekel et al. 1999).
Unfortunately, as distance measurements increase with distance, 
the scales probed are smaller than the interesting 
scale of transition to homogeneity.
Conflicting results 
on both the amplitude  and coherence of the flow suggest that
peculiar velocities cannot yet set strong constraints on the amplitude
of fluctuations on scales of hundreds of Mpc's.
Perhaps the most promising method for the future is the 
kinematic Sunyaev-Zeldovich effect which allows one to measure the 
peculiar velocities of clusters out to high redshift.

The agreement between the CMB dipole and the dipole anisotropy 
of relatively nearby galaxies argues in favour of large scale 
homogeneity.
The IRAS dipole (Strauss et al 1992, Webster et al 1998, 
Schmoldt et al 1999)
shows an apparent convergence 
of the dipole, with 
misalignment angle of only $15^\circ$.
Schmoldt et al. (1999) claim that 
2/3  of the dipole arises from within a $40 \Mpc$, 
but again it is difficult
to `prove' convergence from catalogues of finite depth.

 \subsection{Radio Sources}

Radio sources in surveys have typical median redshift
${\bar z} \sim 1$, and hence are useful probes of clustering at high
redshift. 
Unfortunately, it is difficult to obtain distance information from
these surveys: the radio luminosity function is very broad, and it is
difficult to measure optical redshifts of distant radio sources.
Earlier studies
claimed that  the distribution of radio sources supports the 
`Cosmological Principle'.
However, 
the wide range in intrinsic luminosities of radio sources
would dilute any clustering when projected on the sky.  
Recent analyses  of
new deep radio surveys (e.g. FIRST)
suggest that radio sources are actually  clustered at least as strongly
as local optical
galaxies 
(e.g. Cress et al. 1996; Magliocchetti et al. 1998).
Nevertheless, on the very large scales the distribution of radio sources
seems nearly isotropic. 
Comparison of the measured quadrupole in a radio sample 
in the Green Bank and Parkes-MIT-NRAO
4.85 GHz surveys 
to the theoretically predicted ones (Baleisis et al. 1998)
offers a crude estimate of the fluctuations on scales $ \lambda \sim 600
h^{-1}$ Mpc.  The derived amplitudes are shown in Figure 3 for the two 
assumed Cold Dark Matter (CDM) models.
Given the problems of catalogue matching and shot-noise, these points should be
interpreted at best as `upper limits', not as detections.

\begin{figure}
\protect\centerline{ 
\psfig{figure=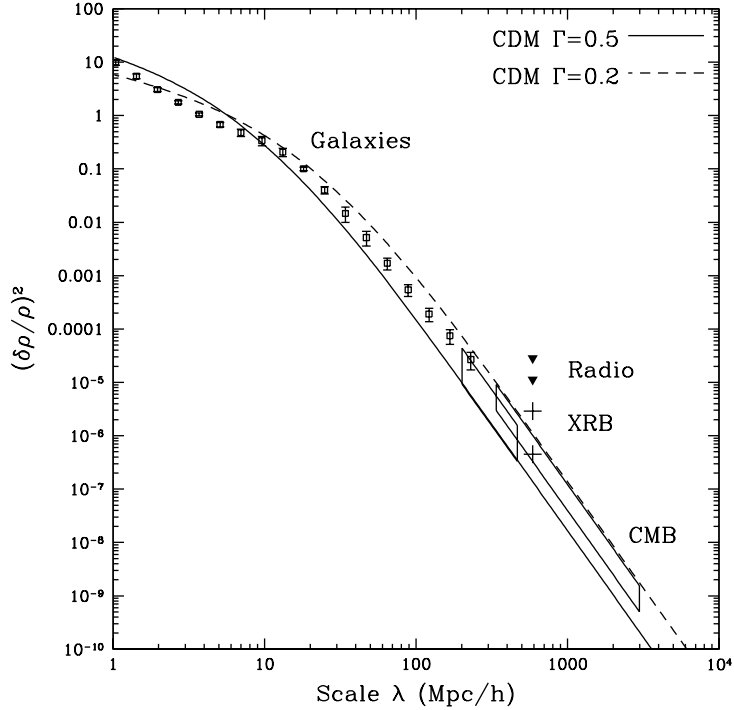,height=4truein,width=4truein}}
\caption[]{
  A compilation of density fluctuations on different scales from
  various observations: a galaxy survey, deep radio surveys, the X-ray
  Background and Cosmic Microwave Background experiments. The
  measurements are compared with two popular Cold Dark Matter models
  (with normalization $\sigma_8=1$ and 
  shape parameters $\Gamma=0.2$ and $0.5$). 
  The Figure shows mean-square density fluctuations $({ {\delta \rho}
    \over \rho })^2 \propto k^3 P(k)$, where $k=1/\lambda$ is the
  wavenumber and $P(k)$ is the power-spectrum of fluctuations.  
  The open squares at
  small scales are estimates from 
  the APM galaxy catalogue (Baugh \& Efstathiou 1994).
  The elongated 'boxes' at large scales represent the COBE
  4-yr (on the right) and
  Tenerife (on the left) CMB measurements  (Gawiser \& Silk
  1998).  The solid triangles and crosses represent amplitudes derived
  from the
  quadrupole of radio sources 
  (Baleisis et al. 1998) and the quadrupole of the XRB 
  (Lahav et al. 1997; Treyer et al. 1998).  
  Each pair of estimates corresponds to assumed shape of the two CDM models.
  (A compilation from Wu, Lahav \& Rees 1999).  }
\end{figure}

 \subsection {The XRB}

 Although discovered in 1962, the origin of
 the X-ray Background (XRB) is still unknown,  
 but is likely
 to be due to sources at high redshift 
 (for review see Boldt 1987; Fabian \& Barcons 1992).
 The XRB sources are probably
 located at redshift $z < 5$, making them convenient tracers of the mass
 distribution on scales intermediate between those in the CMB as probed
 by COBE, and those probed by optical and IRAS redshift
 surveys (see Figure 3).

The interpretation of the results depends somewhat on the nature of
the X-ray sources and their evolution.  
By comparing
the predicted multipoles to those observed by HEAO1 
(Lahav et al. 1997; Treyer et al. 1998; Scharf et al. 2000)
we estimate the amplitude of fluctuations for an
assumed shape of the density fluctuations 
(e.g. CDM models).  
Figure 3 shows the amplitude of fluctuations derived at the 
effective scale $\lambda \sim 600 h^{-1}$ Mpc probed by the XRB. 
The observed fluctuations in the XRB
are roughly as expected from interpolating between the
local galaxy surveys and the COBE CMB experiment.
The rms fluctuations 
${ {\delta \rho} \over {\rho} }$
on a scale of $\sim 600 h^{-1}$Mpc 
are less than 0.2 \%.

\subsection {The Lyman-$\alpha$ Forest}


The Lyman-$\alpha$ 
forest reflects the neutral hydrogen distribution and therefore
is likely to be a more direct  trace of the mass distribution 
than galaxies are.
Unlike galaxy surveys which are
limited to the low redshift Universe, the forest spans a large
redshift interval, typically $1.8 < z < 4$, corresponding 
to comoving interval of $\sim 600 \Mpc$.
Also, observations of the
forest are not contaminated by complex selection effects such as those
inherent in galaxy surveys.  It has been suggested qualitatively by
Davis (1997) that the absence of big voids in the distribution of 
Lyman-$\alpha$
absorbers is inconsistent with the fractal model.
Furthermore, all lines-of-sight towards quasars look
statistically similar.  
Nusser \& Lahav (2000) 
predicted the distribution of the flux  in Lyman-$\alpha$ 
 observations in a specific
truncated fractal-like model. They found that indeed in this model there are
too many voids compared with the observations and conventional (CDM-like)
models
for structure formation.
This too supports the common view that on large scales the Universe 
is homogeneous.
Another test for isotropy, based on the distribution of Supernovae Ia
out to redshift $z \approx 1$ is described in Kolatt \& Lahav (2001).

\section {Is the Universe a Fractal ?}

The question of whether the Universe 
is isotropic and homogeneous on large scales
can also be  phrased in terms of the fractal structure of the 
Universe.
A fractal is a geometric shape that is not homogeneous, 
yet preserves the property that each part is a reduced-scale
version of the whole.
If the matter in the Universe were actually 
distributed like a pure fractal on all scales then the 
Cosmological Principle 
would be invalid, and the standard model in trouble.
As shown in Figure 3
current data already strongly constrain any non-uniformities in the 
galaxy distribution (as well as the overall mass distribution) 
on scales $> 300 \Mpc$.

If we count, for each galaxy,
the number of galaxies within a distance $R$ from it, and call the
average number obtained $N(<R)$, then the distribution is said to be a
fractal of correlation dimension $D_2$ 
if $N(<R)\propto R^{D_2}$. Of course $D_2$
may be 3, in which case the distribution is homogeneous rather than
fractal.  In the pure fractal model this power law holds for all
scales of $R$.

The fractal proponents (Pietronero et al. 1997)  have
estimated $D_2\approx 2$ for all scales up to $\sim 500\Mpc$, whereas
other
groups 
have obtained scale-dependent values 
(for review see Wu et al. 1999 and references therein).

Estimates of $D_2$ 
from the CMB and the XRB
are consistent with $D_2=3$ to within
$10^{-4}$  on the very
large scales (Peebles 1993; Wu et al. 1999).
While we reject the pure fractal model in this review, the performance
of CDM-like models of fluctuations on large scales have yet to be
tested without assuming homogeneity {\it a priori}. On scales below,
say, $30 \Mpc$, the fractal nature of clustering implies that one has
to exercise caution when using statistical methods which assume
homogeneity (e.g. in deriving cosmological parameters).  
We emphasize that we only considered
one `alternative' here, which is the pure fractal model where $D_2$ is a
constant on all scales.

\section {More Realistic Inhomogeneous Models} 

As the Universe appears clumpy on small scales it is clear that 
assuming the Cosmological Principle and the FRW metric is only an 
approximation, and one has to 
average carefully the density in  
Newtonian Cosmology (Buchert \& Ehlers 1997).
Several models in which the matter in clumpy 
(e.g. 'Swiss cheese' and voids)
have been proposed 
(e.g. Zeldovich 1964; Krasinski 1997; Kantowski 1998; Dyer \& Roeder  
1973; Holz \& Wald 1998;  C\'el\'erier 1999; Tomita 1999). 
For example, if the line-of-sight to a distant  object is `empty' 
it results in a gravitational lensing de-magnification of the object.
This modifies the  FRW luminosity-distance relation, with 
a clumping factor as  another free parameter. 
When applied to a sample of SNe Ia
the density parameter  of the Universe
$\omegam$ could be underestimated if FRW is used 
(Kantowski 1998; Perlmutter et al. 1999).
Metcalf \& Silk (1999) pointed out that this effect can be used as a test 
for the nature of the dark matter, i.e. to test if it is smooth  
or clumpy.

\section {Cosmological Parameters from a Joint Analysis: a Cosmic Harmony ? }

A simultaneous analysis of the constraints placed on  cosmological
parameters by different kinds of data is essential because
each probe (e.g. CMB,  
SNe Ia, redshift surveys, cluster abundance
and peculiar velocities)
typically constrains a different combination of
parameters. By performing joint likelihood analyses, one can
overcome intrinsic degeneracies inherent in any single analysis
and so estimate fundamental
parameters much more accurately. The comparison of
constraints can also provide a test for the validity of the assumed
cosmological model or, alternatively, a revised evaluation of the
systematic errors in one or all of the data sets.  Recent papers that
combine information from several data sets simultaneously include
Webster et al. (1998); Lineweaver (1998); 
Gawiser \& Silk (1998),  Bridle et al. (1999, 2001), 
Eisenstein, Hu \& Tegmark 1999; Efstathiou et al. 1999;  
and Bahcall et
al. (1999). 

While joint Likelihood analyses employing both CMB and LSS data are
allowing more accurate estimates of cosmological
parameters, they involve various subtle statistical issues:
\begin{itemize}

\item There is the uncertainty that a sample does not represent 
a typical patch of the FRW Universe to yield reliable global cosmological 
parameters.
\item The choice of the model parameter space is somewhat arbitrary.
\item One commonly solves for the probability for the data given a model
      (e.g. using a Likelihood function),  
      while in the Bayesian framework this should be modified
      by the prior for the model and its parameters.
\item If one is interested in a small set of parameters, should one marginalise
      over all the remaining parameters, rather than  fix them at certain 
      (somewhat ad-hoc) values ?  
\item The `topology' of the Likelihood contours may not be simple. 
      It is helpful when the Likelihood contours of different probes 
      `cross' each other to yield a global maximum 
       (e.g. in the case of CMB and SNe), but in other cases
       they may yield distinct separate `mountains', and the joint 
       maximum Likelihood may lie in a `valley'.
\item Different probes might be spatially correlated, i.e. 
       not necessarily independent.
\item What weight should one give to each data set ?
\end{itemize}

In a long term collaboration in Cambridge (Bridle
et al. 1999, 2001; Efstathiou et al. 1999; Lahav et al. 2000) we have
compared and combined in a self-consistent way the most powerful
cosmic probes: CMB, galaxy redshift surveys, galaxy cluster number
counts, type Ia Supernovae and galaxy peculiar velocities.  These
analyses illustrate the power of combining different data sets for
constraining the fundamental parameters of the Universe.  Our analysis
suggests, in agreement with studies by other groups, that we live in a
flat accelerating Universe, with comparable amounts of dark matter and
`vacuum energy' (cosmological constant). 
We have also addressed recently (Lahav et al. 2000; Lahav 2001) the issue of
combining different data sets, which may suffer different systematic
and random errors.  We generalised the standard procedure of combining
likelihood functions by allowing freedom in the relative weights of
various probes.  This is done by including in the joint likelihood
function a set of `Hyper-Parameters', which are dealt with using
Bayesian considerations.  The resulting algorithm, which assumes
uniform priors on the logarithm of the Hyper-Parameters, is simple to
implement.
Here we show some examples of and results from the joint analysis.
First combining two CMB data sets, and then combining CMB, Supernovae Ia
and peculiar velocities.

\section {Combining the  CMB Boomerang and Maxima Data}

The recent Boomerang (hereafter B; de Bernardis et al. 2000) 
and Maxima (hereafter M; Hanany et al. 2000) CMB anisotropy measurements 
yielded  high-quality angular power spectra $C_l$ over the 
spherical harmonics $ 400 \simlt  l \simlt  800$.
An important factor in interpreting the data is the calibration error.
The experimental papers quote calibration errors of 10\% and 4\%
(1-sigma in $\Delta T/T$) for B and M, respectively.
The measurements (with B data corrected upward by 10\%, 
and M data corrected downward by 4 \%) 
are shown in Figure 4, and they indicate a  
well defined  first acoustic peak at $l \sim 200$, with less convincing 
second and third peaks at higher harmonics.
These measurements favour (under certain assumptions) 
a flat universe, spectral index $n=1$ 
and baryon density $\omegab h^2 \sim 0.03$ 
(e.g. Jaffe et al. 2000), 
which is about 2-sigma
higher than the Big-Bang Nucleosynthesis (BBN) value 
$\omegab h^2 \sim 0.0190 \pm 0.0018$ (95 \% CL; Burles et al. 2000).
Note that the recent CBI result (Padin et al. 2000) gives a higher
power (at $l\sim 600$) relative to  B\&M.
Jaffe et al. (2000) fitted models after 
combining the B\& M data sets into one set. 
We have  taken a different approach for joint analysis of these two data sets 
by utilising the `Hyper-Parameters' (Lahav 2001).

\begin{figure}
\protect\centerline{
\psfig{figure=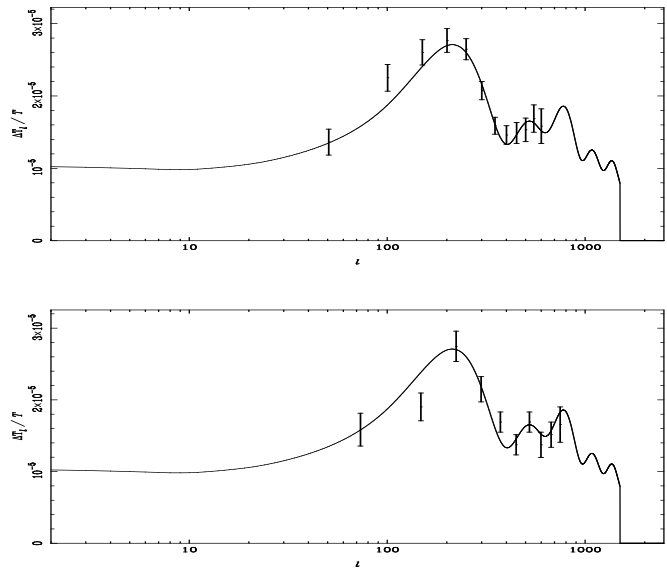,height=5.0truein,width=6.0 truein}}
\caption[]{
The Boomerang $ \Delta T \over T$ data against spherical harmonic $l$
(top panel,
calibrated by 1.10) and Maxima data (bottom panel, calibrated by
0.96). The line in each panel is for a $\Lambda$-CDM model with $n=1,
\omegam = 1 -\lambda = 0.3, \omegab h^2 = 0.03, \qrms = 18 \mu K$,
and our `best fit' $h = 0.8$. }  
\end{figure}

For simplicity, we restrict ourselves to a very limited
set of cosmological models. 
We obtain theoretical CMB power-spectra using the CMBFAST and 
CAMB codes (Slejak \& Zaldarriaga 1996; Lewis, Challinor \& Lasenby 2000).
We assume that CMB fluctuations arise from
adiabatic initial conditions with Cold Dark Matter (CDM) and negligible
tensor component, in a flat Universe with  $\omegam=0.3$,
$\lambda \equiv \Lambda/(3 H_0^2) = 1-\omegam=0.7$, $n=1$,
$\qrms=18\mu$K and $\omegab h^2 = 0.03$.
This choice is motivated by numerous other studies which combined
CMB data with other cosmological probes (e.g. Jaffe et al. 2000,
Bridle et al. 2000; Hu et al. 2000; see also the next section).  
We then
investigate the constraints on the remaining parameter, the
dimensionless Hubble constant, $h=H_0/$(100 km/sec/Mpc). 
Increasing
$h$ decreases the height of the first acoustic peak, and makes few
other significant changes to the angular power spectrum (e.g. Hu et al. 2000). 
The range
in $h$ investigated here is ($0.5<h<1.1$).

The calibration of the data 
brings the two data
sets to much better agreement.
In fact, in this case the standard joint $\chi^2$ and the Hyper Parameters 
give the same result, $h=0.79$, with slightly 
smaller error bars in the HPs case ($\pm 0.04;  95\%$ CL).
We also tried the BBN value 
$\omegab h^2 = 0.019$ 
which gives
gives much poorer $\chi^2$ than the value
$\omegab h^2 = 0.03$  
(e.g.  Jaffe et al. 2000 and others).

The best fit Hubble constant is $H_0 = 79 \pm 4$ km/sec/Mpc 
(95\% CL, random errors only)
for a fixed flat CDM $\omegam = 1 - \lambda = 0.3$ model with
$n=1$,
$\qrms=18\mu$K and $\omegab h^2 = 0.03$.
We note that if more cosmological parameters are left free and then 
marginalised over, the error in $h$ would typically be larger.

This combination of $\omegam$ and $H_0$ corresponds gives for 
the age of the Universe  11.9 Gyr.
Our derived $H_0$ is slightly higher 
but still
 consistent with the `final result' of $H_0$ from Cepheids
and other distance indicators 
(Freedman et al. 2000)
$H_0  = 72 \pm (3)_r \pm (7)_s$ km/sec/Mpc 
(1-sigma random and systematic errors).

\section {Combining CMB, Supernovae Ia and Peculiar Velocities} 

A recent study (Bridle et al. 2001) is an example of 
combining 3 different data sets.
We compared and combined likelihood functions for the matter density
parameter $\omegam$, the Hubble constant 
$h$, 
and the normalization $\sigma_8$
(in terms of the variance in the mass density field measured in an
$8 h^{-1}$ Mpc radius sphere) 
from peculiar velocities, CMB (including the Boomerang and Maxima data)
and type Ia
Supernovae. These three data sets directly probe the mass in the
Universe, without the need to relate the galaxy distribution to the
underlying mass via a ``biasing'' relation.  

Our analysis assumes a flat $\Lambda$CDM cosmology with a
scale-invariant adiabatic initial power spectrum and baryonic fraction
as inferred from big-bang nucleosynthesis.  We find that all three
data sets agree well, overlapping significantly at the 2$\sigma$
level. This therefore justifies a joint analysis, in which we find a
best fit model and $95\%$ confidence limits of $\omegam =0.28\, (0.17,0.39)$, $h=0.74\, (0.64,0.86)$, and $\sigma_8=1.17\,
(0.98,1.37)$. In terms of the natural parameter combinations for these
data $\sigma_8\omegam^{0.6}=0.54\, (0.40,0.73)$, 
$\omegam h = 0.21\, (0.16,0.27)$. Also for the best fit point,
$Q_{\rm{rms-ps}} = 19.7\mu$K and the age of the Universe is $13.0$
Gyr. Figure 5 illustrates these results.

This is quite in agreement from results form cluster abundance 
$\omegam^{0.5} \sigma_8 = 0.5 \pm 0.1 $.
(Eke et al. 1998).
By combining the abundance of clusters with the CMB and IRAS 
Bridle et al. (1999) found 
$\omegam= 1 - \lambda
= 0.36$, $h = 0.54$, $\sigma_8 = 0.74$, and
IRAS biasing parameter
$b_{iras} = 1.08$ (with error bars similar to those  above).

\begin{figure}
\vskip 2.5 truein
\includegraphics{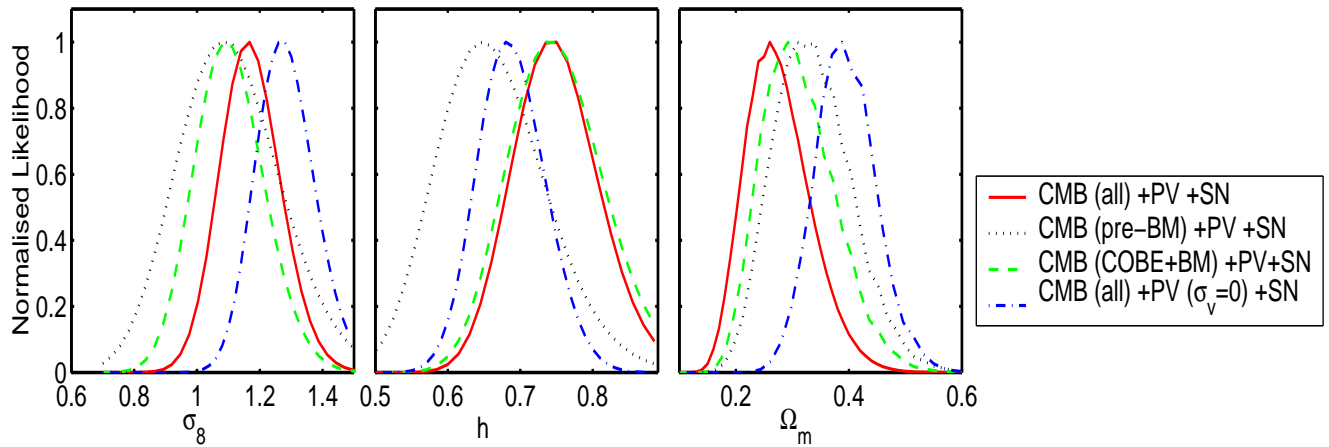}
\caption{
The 1-dimensional marginalised likelihood distributions 
from the joint PV, CMB and SN likelihood function.
The main results are shown by the solid lines, which use the whole CMB
data compilation, PV 
(marginalised over the field velocity dispersion $\sigma_{\rm v}$) and SN. The 
dotted
lines show the likelihood functions when PV, SNe and just the pre-BM
CMB data are used. The dashed line is the result when the CMB data is
just COBE + BOOMERANG + MAXIMA-1. The result (using all the CMB data) 
when the uncorrelated
velocity dispersion term is not included in the PV analysis
($\sigma_{\rm v}=0$) is shown by the dot-dashed line.
(From Bridle etal. 2001.)
    } 
\label{joint}
\end{figure}

\section{Discussion}

Analysis of the CMB, the XRB, radio sources and the Lyman-$\alpha$ 
which probe scales of 
$\sim 100-1000 \Mpc$ strongly support the Cosmological Principle
of homogeneity and isotropy.
They rule out  a pure fractal model.
However, there is a need for more realistic  inhomogeneous models
for the  small scales. This is in particular important for
understanding the validity of cosmological parameters obtained
within the standard FRW cosmology.

Joint analyses of the CMB, IRAS, SNe, cluster abundance and
peculiar velocities suggests   $\omegam=1-\lambda \approx 0.3 $.
The measurement of the Hubble constant from Cepheids 
and from the CMB suggests $H_0 \approx 75 $ km/sec/Mpc.
While this is now a popular model there are potential 
problems with this set of parameters.
There is no simple theoretical explanation why the present epoch 
contributions to matter $\omegam$ and 'dark energy' ($\lambda$) 
are nearly equal.
Also, for the above model the age on the Universe is $\sim 13 $ Gyr
which is uncomfortably close to some estimates for the 
age of the Globular Clusters. 

These issues will not doubt 
be revisited soon with larger and more accurate data sets.
We will soon be able to map
the fluctuations with scale and epoch, and to analyze jointly
redshift surveys 
(2dF, SDSS) and
CMB (MAP, Planck) data. 
These high quality data sets 
will allow us to study a wider range of models and parameters.

\section*{Acknowledgments}

I thank my collaborators for their contribution to the work
presented here, and Jose Lemos and the other oraganisers 
for the hospitality in Lisbon.




\begin{thebibliography}{}  

\bibitem[]{}
Bahcall, N.A., Ostriker, J.P., Perlmutter, S. \& Steinhardt, P.J., 1999, 
Science, 284, 148


\bibitem[]{}
Baleisis, A., Lahav, O., Loan, A.J. \& Wall, J.V. 1998, 
MNRAS, 297, 545



\bibitem[]{}
Baugh C.M. \& Efstathiou G. 1994, MNRAS , 267, 323


\bibitem[]{}
de Bernardis, P. et al., 2000, 
Nature, 404, 955

\bibitem[]{} 
Boldt, E. A. 1987,  Phys. Reports, 146, 215



\bibitem[]{}
Bridle, S.L., Eke, V.R., Lahav, O., Lasenby, A.N., Hobson, M.P., Cole, S., 
Frenk, C.S., \& Henry, J.P. 1999, MNRAS, 310, 565   


\bibitem[]{}
Bridle, S.L., Zehavi, I., Dekel, A., Lahav, O., Hobson, M.P. \& Lasenby, A.N., 
2001, MNRAS, 321, 333


\bibitem[]{}
Buchert T \& Ehlers, J. 1997,  A\&A, 320, 1


\bibitem[]{}
Burles, S., Nollett, K.M. \&  Turner, M.S., 2000
(astro-ph/0008495)



\bibitem[]{}
C\'el\'erier, M.N. 1999, submitted to A\&A (astro-ph/9907206)


\bibitem[]{}
Cress C.M., Helfand D.J., Becker R.H., Gregg. M.D. \& White, R.L.
1996,  ApJ,   473, 7 



\bibitem[]{}
Davis, M. 1997, 
{\it Critical Dialogues in Cosmology}, World Scientific, ed. N. Turok, pg. 13.


\bibitem[]{}
Dekel, A. et al.,  1999, ApJ, in press (astro-ph/9812197)




\bibitem[]{}
Dyer, C.C.  \& Roeder, R.C.  1973, ApJ, 180, L31


\bibitem[]{}
Ehlers, J., Geren, P \& Sachs, R.K. 1968, J Math Phys, 9(9), 1344

\bibitem[]{}	
Efstathiou G., Bridle S.~L., Lasenby A.~N., Hobson M.~P. \&
Ellis R.~S. 1999, MNRAS, 303, L47


\bibitem[]{}
Eisenstein, D.J., Hu, W. \& Tegmark, M. 1999, ApJ, 518, 2




\bibitem[]{}
Eke, V.R.,  Cole, S., Frenk, C.S. \&  Henry, J.P. 1998, MNRAS, 298, 1145



\bibitem[]{}
 Fabian, A. C. \& Barcons, X. 1992,  ARAA, 30, 429

\bibitem[]{}
Freedman, W.L., et al., 2000, ApJ, in press (astro-ph/0012376)



\bibitem[]{}
 Gawiser, E. \& Silk, J.,  1998, Science, 280, 1405







\bibitem[]{}
Hanany, S. et al., 2000,
submitted to ApJL (astro-ph/0005123)

\bibitem[]{}
Holz, D.E. \& Wald, R.M. 1998, Phys Rev D, 58, 063501 


\bibitem[]{}
Hu, W., Fukugita, M., Zaldarriaga, M., Tegmark, M., 2000, submitted
to ApJ (astro-ph/0006436)

\bibitem[]{}
Jaffe et al., 2000,
submitted to PRL (astro-ph/0007333)


\bibitem[]{}
Joyce, M., Montuori, M., Sylos-Labini F. \& Pietronero, L., 1999,
A\&A, 344, 387



\bibitem[]{}
Kantowski, R. 1998, ApJ,  507, 483  

\bibitem[]{}
Kolatt, Ts. \& Lahav, O., 2001,
MNRAS, in press (astro-ph/0008041)

\bibitem[]{}
Krasinski, A. 1997, {\it Inhomogeneous Cosmological Models}, 
Cambridge University Press, Cambridge





\bibitem[]{}
 Lahav O., Piran T. \& Treyer M.A.  1997,  MNRAS, 284, 499


\bibitem[]{}
Lahav, O., Santiago, B.X., Webster, A.M., Strauss, M.A., 
Davis, M., Dressler, A. \& Huchra, J.P.  2000, MNRAS, 312, 546 

\bibitem[]{}
Lahav, O., Bridle, S.L., Hobson, M.P., Lasenby, A.L., Sodr\'e, L. 
2000, MNRAS, 315, 45L



\bibitem[]{}
Lahav, O., 
in the proceedings of IAU201 {\it New Cosmological Data 
and the Values of the Fundamental Parameters} 
Manchester 2001, Eds. A. Lasenby and A. Wilkinson,
in press, (astro-ph/0012475)

\bibitem[]{}
Lewis, A., Challinor, A. \&  Lasenby, A. 2000, ApJ, in press 
(astro-ph/9911177)


\bibitem[]{}
Lineweaver, C. H. 1998, ApJ, 505, L69

\bibitem[]{}
 Magliocchetti, M.,  Maddox, S.J., Lahav, O.\&  Wall, J.V. 1998, 
MNRAS, 300, 257


\bibitem[]{}
Metcalf, R. B.  \&  Silk, J. 1999, ApJ L, 519,  L1


\bibitem[]{}
Nusser, A. \& Lahav, O. 2000, MNRAS, 313, 39L

\bibitem[]{}
Padin, S. et al, submitted to ApJ (astro-ph/0012211)


\bibitem[]{}
Peebles, P. J. E. 1993, {\it  Principles of Physical Cosmology},
Princeton University Press, Princeton.

\bibitem[]{}
Peacock, J.A., et al., 2001, Nature, 410, 169



\bibitem[]{}
Perlmutter et al. 1999, ApJ, 517, 565



\bibitem[]{}
 Pietronero, L., Montuori M., \& Sylos-Labini, F. 1997, in 
{\it Critical Dialogues in Cosmology}, World Scientific, ed. N. Turok, pg. 24



\bibitem[]{}
Rudnicki, K. 1995, 
{\it The cosmological principles}, Jagiellonian University, 
Krakow 


\bibitem[]{}
Scaramella, R. et al. 1998, A\&A, 334, 404


\bibitem[]{}
Scharf, C.A.,
Jahoda, K.,  Treyer, M., Lahav, O., Boldt, E. \&  Piran, T.,  
, 2000, ApJ,  544, 49


\bibitem[]{}
Schmoldt, I. et al. 1999, MNRAS, 304, 893


\bibitem[]{}
Seljak, U. \&  Zaldarriaga, M. 1996, ApJ, 469, 437

\bibitem[]{}
Strauss M.A. et al., 1992, ApJ, 397, 395 




\bibitem[]{}
Tomita, K. 1999 (astro-ph/9906027)



\bibitem[]{}
 Treyer, M., Scharf, C., Lahav, O., 
Jahoda, K.,  Boldt, E. \& Piran, T. 1998,  
ApJ, 509, 531


\bibitem[]{}
Webster, M.A., Lahav, O., \& Fisher, K.B. 1998, MNRAS, 287, 425



\bibitem[]{}
Webster, M., Bridle, S.L., Hobson, M.P., Lasenby, A.N., 
Lahav, O.,  \& Rocha, G. 1998, 
ApJ, 509, L65



\bibitem[]{}
Wu, K.K.S.,  Lahav, O. \&   Rees, M.J. 1998,  Nature, 
397, 225









\bibitem[]{}
Zeldovich, Ya, B. 1964, Soviet Astron, 8, 13

\end{thebibliography}
\end{document}